\documentclass[journal]{IEEEtran}
\usepackage{amsmath,amsfonts,amssymb}
\usepackage{algorithmic}
\usepackage{algorithm}
\usepackage{array}
\usepackage[caption=false,font=normalsize,labelfont=sf,textfont=sf]{subfig}
\usepackage{textcomp}
\usepackage{stfloats}
\usepackage{url}
\usepackage{verbatim}
\usepackage{graphicx}
\usepackage{cite}
\usepackage{tikz}
\usetikzlibrary{positioning,arrows.meta,calc}

\begin{document}
	
	\title{Robust Adaptive Channel Estimation for RIS-Assisted mmWave Massive MIMO via a Stochastic Shannon Entropy Criterion}
	
	%Robust Sparse Channel Estimation for RIS-Assisted mmWave Massive MIMO via Kernel-Based Adaptive Learning
	
	\author{Mojtaba Hajiabadi, Saeid Pakravan, Hadi Zayyani, Mohammad Salman, and Ming Zeng
    
		\thanks{M. Hajiabadi is with the Department of Electrical \& Computer Engineering, University of Birjand, Birjand, Iran (e-mail: mhajiabadi@birjand.ac.ir).}
		\thanks{S. Pakravan is with the Department of Computer Science, University of Quebec in Montreal (UQAM), Montreal, QC, Canada (e-mail: pakravan.saeid@uqam.ca).}
		\thanks{H. Zayyani is with the Department of Electrical and Computer Engineering, Qom University of Technology, Qom, Iran (e-mail: zayyani@qut.ac.ir).}
		\thanks{M. Salman is with the College of Engineering and Technology, American University of the Middle East, Egaila 54200, Kuwait (e-mail: mohammad.salman@aum.edu.kw).}

	\thanks{M. Zeng is with the Department of Electrical and Computer Engineering, Laval University, Quebec City, QC, Canada. (e-mail: ming.zeng@gel.ulaval.ca).}
	
	}

	\maketitle

	\begin{abstract}
		Accurate channel estimation in reconfigurable intelligent surface (RIS)-assisted millimeter-wave (mmWave) massive multiple-input multiple-output (MIMO) systems remains challenging due to the passive nature of the RIS, the high dimensionality of the cascaded channel, and the presence of non-Gaussian impulsive disturbances. This paper proposes a robust sparsity-aware adaptive estimator, termed reweighted zero-attracting stochastic Shannon entropy (RZA-SSE), for angular-domain channel estimation. The algorithm combines a double-window kernel-based Shannon entropy criterion with a norm-normalized \(\ell_1\) penalty. Both the data-driven term and the nonsmooth penalty are differentiated using exact Wirtinger calculus. An exact weight-error recursion and a mean-square energy relation are derived. Local stability is analyzed via a joint coefficient--penalty state representation, and an approximate steady-state mean-square deviation (MSD) expression is obtained. Extensive simulations under Gaussian-mixture impulsive noise demonstrate that ZA-SSE consistently outperforms the state-of-the-art benchmarks in steady-state MSD.
	\end{abstract}


	\section{Introduction}
	
	Millimeter-wave (mmWave) communications, combined with massive
	multiple-input multiple-output (MIMO) technology, offer large
	bandwidth and substantial spatial multiplexing capabilities
	for high-data-rate wireless systems
	\cite{hemadeh2018millimeter,busari2017millimeter}.
	However, mmWave links are particularly vulnerable to
	blockage and propagation attenuation, making accurate
	channel state information essential for effective
	beamforming and precoding \cite{he2020cascaded}.
	Reconfigurable intelligent surfaces (RISs) enable
	programmable control of signal reflections and can
	provide alternative propagation paths when direct
	links are weak or obstructed
	\cite{wu2020towards,di2020smart,wu2020intelligent}.
	
	Reliable channel estimation (CE) in RIS-assisted mmWave
	massive MIMO systems remains challenging. Conventional
	passive RIS elements lack dedicated radio-frequency
	chains, generally precluding direct channel
	measurements at the surface. Moreover, the large
	number of reflecting elements increases the
	dimensionality of cascaded CE
	\cite{wang2020compressed}.
	Owing to the limited number of dominant propagation
	paths, the constituent mmWave channels often exhibit
	sparse angular-domain structure, which can induce
	structured or approximate sparsity in their cascaded
	representation \cite{vuokko2007measurement}.
	Although this structure can facilitate channel
	recovery, estimation accuracy may deteriorate when
	training observations are corrupted by non-Gaussian
	impulsive noise \cite{ashraf2023effect}.
	
	Sparsity-exploiting approaches have been investigated
	for RIS-assisted channel recovery, including structured
	and adaptive estimation methods
	\cite{shukla2024adaptive,shukla2023cascaded}.
	Adaptive filtering offers an iterative framework for
	processing sequential training observations
	\cite{wang2016improved}.
	Conventional least mean squares (LMS) and normalized
	LMS (NLMS) algorithms rely on squared-error adaptation
	\cite{elangovan2012comparative}.
	Sparsity-aware extensions, including zero-attracting
	(ZA) and reweighted zero-attracting (RZA) adaptive
	algorithms, incorporate regularization terms that
	promote sparse coefficient estimates
	\cite{bishnu2017sparse}.
	Reweighted regularization applies weaker attraction
	to coefficients with larger magnitudes, helping
	preserve dominant channel components. However,
	sparsity regularization alone does not address the
	sensitivity of squared-error adaptation to
	impulsive disturbances.
	
	Robust adaptive filtering has therefore been
	investigated using nonlinear error criteria.
	The maximum correntropy criterion (MCC) and its
	complex-valued extensions provide alternatives
	to conventional squared-error minimization
	under non-Gaussian disturbances
	\cite{guimaraes2019maximum,qian2020generalized}.
	Information-theoretic learning also includes
	minimum error entropy (MEE) approaches, which
	exploit kernel-based estimates of error-distribution
	properties rather than relying exclusively
	on second-order statistics
	\cite{principe2010information}.
	In particular, the stochastic information
	gradient framework employs a logarithmic
	criterion based on kernel density estimates
	evaluated over finite observation windows
	\cite{erdogmus2003online}.
	Shannon-entropy-based learning has also been
	investigated for complex-valued adaptive
	filtering \cite{qian2021complex}.
	These developments establish the foundations
	for stochastic error-distribution-based
	adaptation, while the resulting estimation
	behavior depends on the adopted criterion,
	kernel parameters, and statistical properties
	of the observations.
	
	Recent studies have addressed adaptive
	estimation specifically in RIS-assisted
	wireless systems. In \cite{shao2025adaptive},
	the NLMS-LOG and NLMS-NLMF-LOG algorithms
	were developed for sparsity-aware adaptive
	CE in RIS-assisted mmWave
	massive MIMO systems. The former incorporates
	logarithmic sparsity regularization, whereas
	the latter additionally employs a hybrid
	error-based adaptation strategy.
	Correntropy-based stochastic gradient
	adaptation has also been investigated
	for RIS-assisted CE under
	impulsive noise \cite{hajiabadi}.
	These studies motivate a joint treatment of impulsive-noise robustness and angular-domain sparsity in complex-valued adaptive CE.

	In this paper, we develop a robust
	sparsity-aware adaptive channel estimator,
	termed reweighted zero-attracting
	stochastic Shannon entropy (RZA-SSE),
	for RIS-assisted mmWave massive MIMO
	systems. Building on stochastic
	kernel-based entropy learning, the
	proposed formulation combines a
	Shannon-entropy-inspired logarithmic
	criterion, constructed from Gaussian-kernel
	evaluations of complex error differences,
	with a saturating sparsity-promoting
	penalty. Differentiating the logarithmic
	criterion yields an adaptive update
	involving normalized kernel weights
	determined by error differences within
	a finite observation window. The
	regularization term imposes progressively
	weaker shrinkage on coefficients of
	larger magnitude, promoting sparsity
	while reducing the direct regularization
	applied to dominant channel coefficients.
	The theoretical investigation establishes
	exact weight-error and energy relations
	and examines local stability and
	steady-state behavior under explicitly
	stated statistical and linearization
	assumptions.
	
	The main contributions of this paper
	are summarized as follows:
	
	\begin{itemize}
		
		\item A sparsity-aware RZA-SSE estimator
		is formulated for angular-domain CE
		in RIS-assisted mmWave massive MIMO
		systems. The proposed objective integrates
		a stochastic kernel-based logarithmic
		error criterion with a saturating,
		magnitude-dependent sparsity penalty.
		The corresponding complex-valued
		adaptive update is derived using
		Wirtinger calculus.
		
		\item An exact weight-error recursion
		and a mean-square energy identity
		are derived for the proposed algorithm.
		Its mean behavior and local stability
		are examined, and an approximate
		steady-state mean-square deviation
		(MSD) relation is obtained through
		local linearization under stated
		statistical assumptions.
		
		\item The proposed estimator is evaluated
		under Gaussian-mixture impulsive noise
		against representative sparsity-aware
		and robust adaptive estimators.
		Numerical experiments assess convergence
		and steady-state MSD over different
		adaptation step sizes, signal-to-noise
		ratios (SNRs), and kernel bandwidths.
		
	\end{itemize}
	
	The remainder of this paper is organized
	as follows. Section II presents the
	RIS-assisted mmWave massive MIMO system
	and observation models. Section III
	develops the RZA-SSE criterion and
	adaptive update. Section IV analyzes
	the weight-error dynamics, local
	stability, and steady-state performance.
	Section V presents the simulation
	results, and Section VI concludes
	the paper.
	
	\textit{Notation:}
	Bold uppercase and lowercase letters
	denote matrices and vectors, respectively.
	The superscripts $(\cdot)^T$, $(\cdot)^H$,
	and $(\cdot)^*$ denote transpose,
	Hermitian transpose, and complex
	conjugation, respectively.
	The operator $\operatorname{diag}(\mathbf{x})$
	forms a diagonal matrix from $\mathbf{x}$,
	and $\operatorname{vec}(\cdot)$ denotes
	column-wise vectorization.
	The symbols $\|\cdot\|_1$ and $\|\cdot\|_2$
	denote the $\ell_1$- and $\ell_2$-norms,
	respectively, $\operatorname{Tr}(\cdot)$
	denotes the matrix trace, and
	$\mathbb{E}[\cdot]$ denotes expectation.
	The symbol $\mathbf{I}$ denotes an
	identity matrix of appropriate dimension.
	For $z\in\mathbb{C}$,
	$\operatorname{sgn}(z)=z/|z|$ when
	$z\neq 0$, with
	$\operatorname{sgn}(0)=0$; for vectors,
	the operator is applied elementwise.

	\section{System Model}
	
	\begin{figure}[!t]
		\centering
		\begin{tikzpicture}[
			scale=0.95,
			font=\footnotesize,
			bs/.style={draw,thick,rounded corners=2pt,fill=blue!12,minimum width=1.5cm,minimum height=2.5cm},
			ris/.style={draw,thick,fill=green!12,minimum width=2.8cm,minimum height=0.55cm},
			ue/.style={draw,thick,circle,fill=orange!25,minimum size=0.75cm},
			arrow/.style={->,thick,>=stealth},
			block/.style={->,thick,dashed,red!70!black,>=stealth}
			]
			% ===== Base Station =====
			\node[bs] (bs) at (0,0) {};
			\node[above=0.12cm of bs.north,font=\small\bfseries] {BS};
			\node[below=0.08cm of bs.south,align=center] {$N_B$ antennas\\(massive array)};
			% BS antenna elements (vertical)
			\foreach \y in {-0.95,-0.55,-0.15,0.25,0.65}
			\draw[fill=blue!45] (-0.52,\y) rectangle (-0.28,\y+0.22);
			
			% ===== RIS (horizontal panel) =====
			\node[ris] (ris) at (4.3,1.4) {};
			\node[above=0.12cm of ris.north,font=\small\bfseries] {RIS};
			\node[below=0.08cm of ris.south,align=center] {$N_R$ passive elements};
			% RIS elements (horizontal)
			\foreach \x in {-1.15,-0.75,-0.35,0.05,0.45,0.85,1.25}
			\draw[fill=green!45] (4.3+\x-0.15,1.25) rectangle (4.3+\x+0.15,1.55);
			
			% ===== User =====
			\node[ue] (ue) at (7.6,-1.5) {};
			\node[below=0.12cm of ue,font=\small\bfseries] {User $k$};
			
			% ===== Paths =====
			% Blocked direct path
			\draw[block] (ue.west) -- (bs.south east);
			\node[red!70!black,font=\scriptsize,rotate=15] at (3.5,-1.45) {blocked};
			
			% User → RIS
			\draw[arrow,blue!70!black] (ue.north) to[out=85,in=-25] (ris.south east);
			\node[blue!70!black,font=\scriptsize] at (6.6,0.15) {$\mathbf{h}_{R,k}$};
			
			% RIS → BS
			\draw[arrow,blue!70!black] (ris.west) to[out=180,in=30] (bs.east);
			\node[blue!70!black,font=\scriptsize] at (2.1,1.55) {$\mathbf{H}_{BR}$};
			
			% Phase-shift annotation
			\node[font=\scriptsize,align=center] at (4.3,0.35) {$\boldsymbol{\theta}_q$\\phase shifts};
			
			% Angular-domain note
			\node[draw,rounded corners,fill=yellow!12,align=left,font=\scriptsize] at (4,-0.7)
			{Angular-domain sparsity:\\$\widetilde{\mathbf{H}}_k=\mathbf{U}_B^H\mathbf{H}_k\mathbf{U}_R$};
		\end{tikzpicture}
		\caption{Uplink RIS-assisted mmWave massive MIMO system. The direct link is blocked; communication occurs through the cascaded channel \(\mathbf{H}_k=\mathbf{H}_{BR}\operatorname{diag}(\mathbf{h}_{R,k})\).}
		\label{fig:system_model}
	\end{figure}
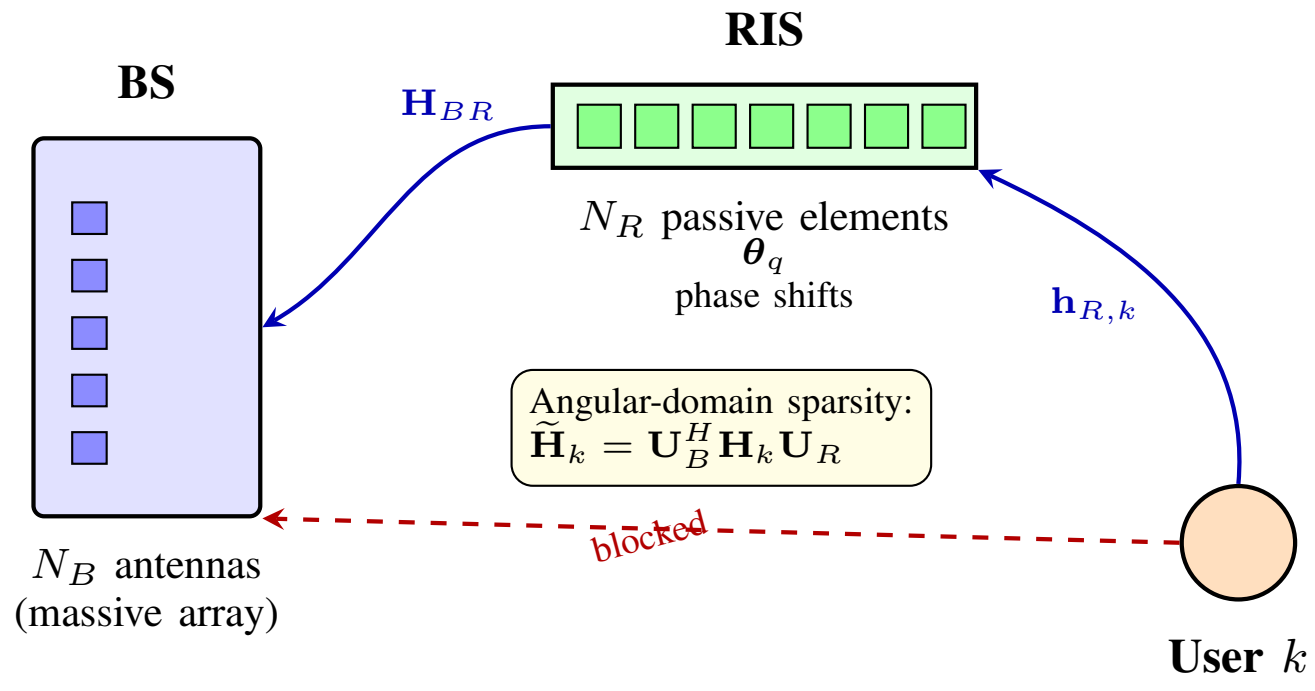
	
	\subsection{Cascaded Channel Model}
	
	As shown in Fig. \ref{fig:system_model}, we consider an uplink RIS-assisted mmWave massive MIMO
	system in which a base station (BS) equipped with
	$N_B$ antennas serves $K$ single-antenna users
	through an RIS comprising $N_R$ passive reflecting
	elements. The direct user--BS links are assumed
	to be blocked, and the propagation channels
	remain approximately constant over each channel
	estimation interval.
	
	Let $\mathbf{H}_{BR}\in\mathbb{C}^{N_B\times N_R}$
	denote the RIS-to-BS channel and
	$\mathbf{h}_{R,k}\in\mathbb{C}^{N_R}$
	the channel from user $k$ to the RIS.
	Following the geometric Saleh--Valenzuela
	model \cite{wang2020compressed}, these channels
	are expressed as
	\begin{equation}
		\mathbf{H}_{BR}
		=
		\sqrt{\frac{N_BN_R}{L_b}}
		\sum_{\ell=1}^{L_b}
		\alpha_{\ell}^{b}
		\mathbf{b}(\vartheta_{\ell}^{r},\psi_{\ell}^{r})
		\mathbf{a}^{H}(\vartheta_{\ell}^{t},\psi_{\ell}^{t}),
		\label{eq:1}
	\end{equation}
	and
	\begin{equation}
		\mathbf{h}_{R,k}
		=
		\sqrt{\frac{N_R}{L_{r,k}}}
		\sum_{\ell=1}^{L_{r,k}}
		\alpha_{r,k}^{\ell}
		\mathbf{a}(\vartheta_{r,k}^{\ell},\psi_{r,k}^{\ell}),
		\label{eq:2}
	\end{equation}
	where $L_b$ and $L_{r,k}$ denote the numbers
	of dominant paths in the RIS--BS and
	user--RIS links, respectively.
	The coefficients $\alpha_{\ell}^{b}$ and
	$\alpha_{r,k}^{\ell}$ are the corresponding
	complex path gains, while the unit-norm
	vectors $\mathbf{b}(\cdot)\in\mathbb{C}^{N_B}$
	and $\mathbf{a}(\cdot)\in\mathbb{C}^{N_R}$
	represent the BS and RIS array responses.
	The angular arguments specify the propagation
	directions according to the adopted array geometry.
	
	For user $k$, the cascaded channel is defined as
	\cite{he2020cascaded}
	\begin{equation}
		\mathbf{H}_k
		=
		\mathbf{H}_{BR}
		\operatorname{diag}(\mathbf{h}_{R,k})
		\in\mathbb{C}^{N_B\times N_R}.
		\label{eq:3}
	\end{equation}
	This representation permits estimation of the
	cascaded channel without separately identifying
	its constituent links. The limited number of
	dominant mmWave propagation paths motivates
	an angular-domain representation. Using unitary
	dictionary matrices
	$\mathbf{U}_B\in\mathbb{C}^{N_B\times N_B}$
	and
	$\mathbf{U}_R\in\mathbb{C}^{N_R\times N_R}$,
	we write \cite{wei2022channel}
	\begin{equation}
		\mathbf{H}_k
		=
		\mathbf{U}_B
		\widetilde{\mathbf{H}}_k
		\mathbf{U}_R^H,
		\label{eq:4}
	\end{equation}
	where
	$\widetilde{\mathbf{H}}_k
	=\mathbf{U}_B^H\mathbf{H}_k\mathbf{U}_R$
	is the angular-domain cascaded channel.
	Although (4) is an exact change of basis,
	$\widetilde{\mathbf{H}}_k$ is generally
	approximately sparse because off-grid
	propagation directions may introduce
	angular leakage.
	
	\subsection{Problem Formulation}
	
	We adopt user-specific training intervals
	in which only the scheduled user transmits
	pilot symbols. For user $k$, the BS collects
	$Q$ observations under known RIS phase
	configurations
	$\boldsymbol{\theta}_q\in\mathbb{C}^{N_R}$,
	$q=1,\ldots,Q$. Under ideal phase-only
	reflection,
	$|[\boldsymbol{\theta}_q]_r|=1$
	for each RIS element $r$.
	The received signal in training slot $q$ is
	\begin{equation}
		\mathbf{y}_{k,q}
		=
		\mathbf{H}_k
		\boldsymbol{\theta}_q s_{k,q}
		+
		\mathbf{v}_{k,q},
		\label{eq:5}
	\end{equation}
	where $s_{k,q}$ is a known pilot symbol and
	$\mathbf{v}_{k,q}\in\mathbb{C}^{N_B}$
	denotes the additive disturbance.
	For notational simplicity, we set
	$s_{k,q}=1$; known nonzero pilot symbols
	can be incorporated into the sensing vectors.
	
	Stacking the observations yields
	\begin{equation}
		\mathbf{Y}_k
		=
		\mathbf{H}_k\boldsymbol{\Theta}
		+
		\mathbf{V}_k,
		\label{eq:6}
	\end{equation}
	where
	$\mathbf{Y}_k,\mathbf{V}_k
	\in\mathbb{C}^{N_B\times Q}$
	and
	$\boldsymbol{\Theta}
	=[\boldsymbol{\theta}_1,\ldots,
	\boldsymbol{\theta}_Q]
	\in\mathbb{C}^{N_R\times Q}$.
	Substituting (4) into (6) and applying
	the BS angular transform gives
	\begin{equation}
		\mathbf{Z}_k
		=
		\widetilde{\mathbf{H}}_k\mathbf{A}
		+
		\mathbf{N}_k,
		\label{eq:7}
	\end{equation}
	where
	$\mathbf{Z}_k=\mathbf{U}_B^H\mathbf{Y}_k$,
	$\mathbf{A}=\mathbf{U}_R^H\boldsymbol{\Theta}$,
	and
	$\mathbf{N}_k=\mathbf{U}_B^H\mathbf{V}_k$.
	
	To obtain a scalar adaptive regression,
	consider the $b$-th row of (7).
	Define
	$\mathbf{g}_{k,b}
	=[\widetilde{\mathbf{H}}_k]_{b,:}^{T}
	\in\mathbb{C}^{N_R}$
	and
	$\mathbf{a}_q=[\mathbf{A}]_{:,q}
	\in\mathbb{C}^{N_R}$.
	The corresponding observation satisfies
	\begin{equation}
		z_{k,b}(q)
		=
		\mathbf{g}_{k,b}^{T}\mathbf{a}_q
		+
		n_{k,b}(q),
		\label{eq:scalar_observation}
	\end{equation}
	where
	$n_{k,b}(q)=[\mathbf{N}_k]_{b,q}$.
	By defining
	$\mathbf{w}_{o,k,b}=\mathbf{g}_{k,b}^{*}$,
	$\mathbf{x}(q)=\mathbf{a}_q$,
	$d(q)=z_{k,b}(q)$, and
	$v(q)=n_{k,b}(q)$,
	we obtain
	\begin{equation}
		d(q)
		=
		\mathbf{w}_{o,k,b}^{H}\mathbf{x}(q)
		+
		v(q).
		\label{eq:adaptive_regression}
	\end{equation}
	
	For a fixed user and BS angular component,
	the subscripts $(k,b)$ are omitted when
	no ambiguity arises. The adaptive estimator
	maintains
	$\mathbf{w}(q)\in\mathbb{C}^{N_R}$
	and forms the a priori error
	\begin{equation}
		e(q)
		=
		d(q)-\mathbf{w}^{H}(q)\mathbf{x}(q).
		\label{eq:adaptive_error}
	\end{equation}
	For the finite-window criterion introduced
	in Section III, past observations are
	reevaluated using the current coefficient
	vector. Specifically, for $i<q$,
	\begin{equation}
		e_i(\mathbf{w}(q))
		=
		d(i)-\mathbf{w}^{H}(q)\mathbf{x}(i).
		\label{eq:window_error}
	\end{equation}
	Thus, the error differences entering the
	criterion are evaluated consistently
	at the same coefficient vector.
	
	After adaptation, the estimated angular
	coefficient vector is recovered as
	$\widehat{\mathbf{g}}_{k,b}
	=\widehat{\mathbf{w}}_{k,b}^{*}$.
	The row estimates form
	$\widehat{\widetilde{\mathbf{H}}}_k$,
	and the spatial-domain estimate follows
	from (4).
	
	The sensing matrix is determined by
	the RIS training configurations and
	angular dictionary. When $Q<N_R$,
	channel recovery relies on suitable
	sparsity and sensing conditions rather
	than sparsity alone. Moreover, because
	the proposed criterion depends on
	differences between prediction errors,
	the diversity of the sensing-vector
	differences is relevant to coefficient
	identifiability.
	
	\subsection{Impulsive Disturbance Model}
	
	Pilot observations may be corrupted
	by non-Gaussian disturbances caused by
	external interference and other impulsive
	impairments \cite{hu2008gaussian,gulati2012statistical}.
	Such disturbances can produce large
	estimation errors and adversely affect
	squared-error-based adaptive algorithms.
	
	We model the additive BS disturbance
	using a complex Gaussian mixture model
	(GMM) \cite{Mohjazi2022Performance}. Specifically,
	$\mathbf{v}_{k,q}\in\mathbb{C}^{N_B}$
	has the probability density function
	\begin{equation}
		p_{\mathbf{V}}(\mathbf{v})
		=
		\sum_{m=1}^{M_G}
		\omega_m
		\mathcal{CN}
		(\mathbf{v};\boldsymbol{\mu}_m,\mathbf{C}_m),
		\label{eq:gmm_pdf}
	\end{equation}
	where $M_G$ is the number of mixture
	components, $\omega_m\geq0$ are the
	mixing probabilities satisfying
	$\sum_{m=1}^{M_G}\omega_m=1$, and
	$\boldsymbol{\mu}_m$ and $\mathbf{C}_m$
	denote the component mean vectors
	and positive-definite covariance
	matrices, respectively. Each Gaussian
	component is assumed proper conditional
	on its mixture index.
	
	Low-probability components with
	relatively large covariance magnitudes
	can represent impulsive disturbances.
	The adopted GMM is not restricted
	to Middleton's Class A model, which
	imposes additional structure on the
	mixture parameters.
	
	Under the unitary BS angular transform,
	$\mathbf{n}_{k,q}
	=\mathbf{U}_B^H\mathbf{v}_{k,q}$
	also follows a complex GMM, with
	component means
	$\mathbf{U}_B^H\boldsymbol{\mu}_m$
	and covariances
	$\mathbf{U}_B^H\mathbf{C}_m\mathbf{U}_B$.
	The scalar disturbance $v(q)$ in
	(\ref{eq:adaptive_regression})
	is the corresponding marginal component
	of the transformed mixture.
	
	The overall disturbance mean is
	\begin{equation}
		\mathbb{E}[\mathbf{v}_{k,q}]
		=
		\sum_{m=1}^{M_G}
		\omega_m\boldsymbol{\mu}_m.
		\label{eq:gmm_mean}
	\end{equation}
	Hence, a zero-mean disturbance requires
	$\sum_{m=1}^{M_G}
	\omega_m\boldsymbol{\mu}_m=\mathbf{0}$
	and is not implied by the mixture
	representation itself. The temporal
	dependence of the disturbances and
	their statistical relationship with
	the sensing vectors are specified
	as needed for the theoretical analysis
	in Section IV.

	\section{Proposed Algorithm}
	
	This section develops the proposed RZA-SSE algorithm for sparse angular-domain
	CE, based on the complex-valued
	regression model established in Section II.
	
	\subsection{Stochastic Shannon-Entropy Criterion}
	
	For a given user and BS angular component,
	consider the regression model in
	(\ref{eq:adaptive_regression}).
	Suppressing the user and angular-component
	indices, let
	$\mathbf{w}(n)\in\mathbb{C}^{M}$,
	with $M=N_R$, denote the adaptive estimate
	of the unknown coefficient vector
	$\mathbf{w}_o$. The sensing vector
	$\mathbf{x}(n)\in\mathbb{C}^{M}$
	and desired observation $d(n)$
	define the a priori error
	$e(n)=d(n)-\mathbf{w}^{H}(n)\mathbf{x}(n)$,
	as in (\ref{eq:adaptive_error}).
	
	For a complex-valued error random variable
	$E$ with probability density $p_E$ on
	the complex plane, the Shannon
	differential entropy is
	\begin{equation}
		H(E)
		=
		-\mathbb{E}\left[\log p_E(E)\right].
		\label{eq:shannon_entropy}
	\end{equation}
	Direct online optimization of this quantity
	is generally impractical because the
	error distribution is unknown.
	Motivated by stochastic
	information-gradient learning
	\cite{erdogmus2003online},
	we employ a kernel-based logarithmic
	criterion evaluated over a finite
	observation window.
	
	As established in Section II, the
	current and reference errors are
	evaluated using a common coefficient
	vector. For any candidate vector
	$\mathbf{w}$, define
	\begin{equation}
		\Delta e_{ni}(\mathbf{w})
		=
		e_n(\mathbf{w})-e_i(\mathbf{w}),
		\label{eq:error_difference}
	\end{equation}
	where
	$e_i(\mathbf{w})
	=d(i)-\mathbf{w}^{H}\mathbf{x}(i)$.
	At iteration $n$, the preceding
	$L$ observations constitute the
	reference window
	$i=n-L,\ldots,n-1$.
	
	The Gaussian kernel on the complex
	plane is defined as
	\begin{equation}
		K_{\beta}(z)
		=
		\frac{1}{2\pi\beta^2}
		\exp\left(
		-\frac{|z|^2}{2\beta^2}
		\right),
		\qquad \beta>0,
		\label{eq:11}
	\end{equation}
	where $\beta$ is the kernel bandwidth.
	The kernel density estimate evaluated
	at the current error is
	\begin{equation}
		\widehat{p}_{E,n}
		=
		\frac{1}{L}
		\sum_{i=n-L}^{n-1}
		K_{\beta}
		\left(
		\Delta e_{ni}(\mathbf{w}(n))
		\right).
		\label{eq:10}
	\end{equation}
	The proposed stochastic logarithmic
	error criterion is then
	\begin{equation}
		J_{\mathrm{SSE}}(n)
		=
		-\log\left[
		\frac{1}{L}
		\sum_{i=n-L}^{n-1}
		K_{\beta}
		\left(
		\Delta e_{ni}(\mathbf{w}(n))
		\right)
		\right].
		\label{eq:12}
	\end{equation}
	
	Equation (\ref{eq:12}) defines a
	stochastic surrogate inspired by
	Shannon differential entropy rather
	than an exact evaluation of $H(E)$.
	The logarithmic operation introduces
	normalization by the estimated
	error density into the adaptive
	gradient, distinguishing this
	criterion from direct kernel-density
	optimization. Its estimation behavior
	depends on the observation statistics,
	kernel bandwidth, and reference window.
	
	\subsection{RZA-SSE Cost Function and Adaptive Update}
	
	% =========================================================================
	% REVIEWER COMMENT (to be addressed in revision):
	% The penalty is not "RZA" in the Chen-Gu-Hero sense. gamma|w|/(1+eps|w|)
	% gives a 1/(1+eps|w|)^2 attractor that decays like |w|^{-2}. That is closer
	% to an approximate-l0 penalty. The paper should say so and justify the
	% choice. If RZA-MCC [20] uses the log-sum attractor, the comparison mixes
	% two changes at once: the criterion and the penalty.
	%   - Section IV.B states the bound in mu, but eta is the natural parameter.
	%     Give a sufficient condition, e.g. lambda_max(A_n) <= max_i ||Delta x_ni||^2.
	%   - For i.i.d. unit-modulus RIS phases, E[xx^H] = I and A ~ 2I. That gives
	%     a time constant of 1/(2 eta) ~ 300 iterations, consistent with Fig. 1
	%     and worth stating.
	% =========================================================================
	
	To exploit the approximate sparsity
	of the angular-domain cascaded channel,
	we augment the stochastic logarithmic
	criterion with the saturating
	reweighted zero-attraction penalty
	\begin{equation}
		P_{\mathrm{RZA}}(\mathbf{w})
		=
		\gamma
		\sum_{j=1}^{M}
		\frac{|w_j|}
		{1+\epsilon|w_j|},
		\label{eq:14}
	\end{equation}
	where $\gamma>0$ is the regularization
	strength and $\epsilon>0$ controls
	the magnitude-dependent reweighting.
	For $|w_j|\ll1/\epsilon$, each term
	behaves approximately as $|w_j|$,
	whereas for large magnitudes it
	approaches $1/\epsilon$.
	Consequently, the penalty imposes
	weaker marginal shrinkage on dominant
	coefficients. The penalty is nonconvex,
	and global optimality is not guaranteed.
	
	The regularized objective is
	\begin{equation}
		J_{\mathrm{RZA-SSE}}(n)
		=
		J_{\mathrm{SSE}}(n)
		+
		P_{\mathrm{RZA}}(\mathbf{w}(n)).
		\label{eq:19}
	\end{equation}
	The adaptive update follows
	complex-valued stochastic gradient
	descent:
	\begin{equation}
		\mathbf{w}(n+1)
		=
		\mathbf{w}(n)
		-
		\mu
		\nabla_{\mathbf{w}^{*}}
		J_{\mathrm{RZA-SSE}}(n),
		\label{eq:16}
	\end{equation}
	where $\mu>0$ is the step size and
	$\nabla_{\mathbf{w}^{*}}$ denotes
	the Wirtinger gradient with respect
	to the conjugate coefficient vector.

	For notational convenience, let
	$\Delta\mathbf{x}_{ni}=\mathbf{x}(n)-\mathbf{x}(i)$,
	$\Delta e_{ni}=\Delta e_{ni}(\mathbf{w}(n))$,
	and $K_{ni}=K_{\beta}(\Delta e_{ni})$.
	
	The error difference satisfies
	\begin{equation}
		\Delta e_{ni}(\mathbf{w})
		=
		d(n)-d(i)
		-
		\mathbf{w}^{H}\Delta\mathbf{x}_{ni}.
		\label{eq:delta_regression}
	\end{equation}
	Using Wirtinger calculus,
	\begin{equation}
		\nabla_{\mathbf{w}^{*}}
		|\Delta e_{ni}(\mathbf{w})|^2
		=
		-\Delta e_{ni}^{*}(\mathbf{w})
		\Delta\mathbf{x}_{ni}.
		\label{eq:delta_gradient}
	\end{equation}
	Applying the chain rule to
	(\ref{eq:12}) yields
	\begin{equation}
		\nabla_{\mathbf{w}^{*}}
		J_{\mathrm{SSE}}(n)
		=
		-\frac{1}{2\beta^2}
		\frac{
			\displaystyle
			\sum_{i=n-L}^{n-1}
			K_{ni}\Delta e_{ni}^{*}
			\Delta\mathbf{x}_{ni}
		}{
			\displaystyle
			\sum_{i=n-L}^{n-1}K_{ni}
		}.
		\label{eq:17}
	\end{equation}
	
	For $w_j\neq0$, the scalar
	Wirtinger derivative of the
	regularization term is
	\begin{equation}
		\frac{\partial}{\partial w_j^{*}}
		\left(
		\frac{|w_j|}
		{1+\epsilon|w_j|}
		\right)
		=
		\frac{1}{2}
		\frac{\operatorname{sgn}(w_j)}
		{(1+\epsilon|w_j|)^2}.
		\label{eq:rza_scalar_gradient}
	\end{equation}
	Accordingly,
	\begin{equation}
		\nabla_{\mathbf{w}^{*}}
		P_{\mathrm{RZA}}(\mathbf{w})
		=
		\frac{\gamma}{2}
		\frac{
			\operatorname{sgn}(\mathbf{w})
		}{
			(1+\epsilon|\mathbf{w}|)^2
		},
		\label{eq:rza_gradient}
	\end{equation}
	where the operations in the fraction
	are applied elementwise. At zero-valued
	coefficients, the zero element of the
	generalized subdifferential is selected.
	
	Substituting (\ref{eq:17}) and
	(\ref{eq:rza_gradient}) into
	(\ref{eq:16}) gives the RZA-SSE update
	\begin{equation}
		\begin{aligned}
			\mathbf{w}(n+1)
			={}&
			\mathbf{w}(n)
			+
			\eta
			\frac{
				\displaystyle
				\sum_{i=n-L}^{n-1}
				K_{ni}\Delta e_{ni}^{*}
				\Delta\mathbf{x}_{ni}
			}{
				\displaystyle
				\sum_{i=n-L}^{n-1}K_{ni}
			}\\
			&-
			\rho
			\frac{
				\operatorname{sgn}(\mathbf{w}(n))
			}{
				(1+\epsilon|\mathbf{w}(n)|)^2
			},
		\end{aligned}
		\label{eq:20}
	\end{equation}
	where
	$\eta=\frac{\mu}{2\beta^2}
	$ and 
	$\rho=\frac{\mu\gamma}{2}.
	$
	The resulting update combines a
	normalized kernel-weighted error
	gradient with magnitude-dependent
	sparsity regularization.
	
	\subsection{Implementation and Computational Complexity}
	
	For numerical stability, the normalized
	kernel weights are computed as
	\begin{equation}
		\pi_{ni}
		=
		\frac{
			\exp\left(
			-\frac{|\Delta e_{ni}|^2-c_n}
			{2\beta^2}
			\right)
		}{
			\displaystyle
			\sum_{j=n-L}^{n-1}
			\exp\left(
			-\frac{|\Delta e_{nj}|^2-c_n}
			{2\beta^2}
			\right)
		},
		\label{eq:normalized_kernel}
	\end{equation}
	where
	\begin{equation}
		c_n
		=
		\min_{n-L\leq j<n}
		|\Delta e_{nj}|^2.
		\label{eq:kernel_shift}
	\end{equation}
	Subtracting $c_n$ before
	exponentiation leaves the normalized
	weights unchanged while preventing
	simultaneous underflow of all
	kernel terms.
	
	Using these weights, the update in
	(\ref{eq:20}) is equivalently expressed as
	\begin{equation}
		\mathbf{w}(n+1)
		={}
		\mathbf{w}(n)
		+
		\eta
		\sum_{i=n-L}^{n-1}
		\pi_{ni}\Delta e_{ni}^{*}
		\Delta\mathbf{x}_{ni}-
		\rho
		\frac{
			\operatorname{sgn}(\mathbf{w}(n))
		}{
			(1+\epsilon|\mathbf{w}(n)|)^2
		}.
		\label{eq:rza_sse_normalized_update}
	\end{equation}
	The implementation is summarized
	in Algorithm~\ref{alg1}.
	
	\begin{algorithm}[t]
		\caption{Proposed RZA-SSE Adaptive Channel Estimator}
		\label{alg1}
		\begin{algorithmic}[1]
			\STATE \textbf{Input:}
			Sequential observations
			$\{\mathbf{x}(n),d(n)\}$;
			$L$, $\beta$, $\eta$, $\rho$, $\epsilon$
			\STATE Collect the first $L$ observations
			\STATE Initialize $\mathbf{w}(L+1)=\mathbf{0}$
			\FOR{$n=L+1,L+2,\ldots$}
			\STATE Compute
			$e_n=d(n)-\mathbf{w}^{H}(n)\mathbf{x}(n)$
			\FOR{$i=n-L,\ldots,n-1$}
			\STATE Compute
			$e_i=d(i)-\mathbf{w}^{H}(n)\mathbf{x}(i)$
			\STATE Set
			$\Delta e_{ni}=e_n-e_i$
			\STATE Set
			$\Delta\mathbf{x}_{ni}
			=\mathbf{x}(n)-\mathbf{x}(i)$
			\ENDFOR
			\STATE Compute $c_n$ using
			(\ref{eq:kernel_shift})
			\STATE Compute $\pi_{ni}$ using
			(\ref{eq:normalized_kernel})
			\STATE Compute
			$\displaystyle
			\mathbf{g}(n)
			=
			\sum_{i=n-L}^{n-1}
			\pi_{ni}
			\Delta e_{ni}^{*}
			\Delta\mathbf{x}_{ni}$
			\STATE Update
			$\displaystyle
			\mathbf{w}(n+1)
			=
			\mathbf{w}(n)
			+\eta\mathbf{g}(n)
			-\rho
			\frac{
				\operatorname{sgn}(\mathbf{w}(n))
			}{
				(1+\epsilon|\mathbf{w}(n)|)^2
			}$
			\ENDFOR
		\end{algorithmic}
	\end{algorithm}
	
	The normalized kernel weights assign
	greater relative importance to
	reference samples with smaller error
	differences. Nevertheless, kernel
	normalization does not guarantee a
	bounded data gradient or complete
	rejection of impulsive disturbances.
	The resulting estimation behavior
	depends on the error statistics,
	kernel bandwidth, and reference
	window. Furthermore, the nonconvex
	RZA penalty introduces a
	sparsity--bias tradeoff, motivating
	the local stability and steady-state
	analysis presented in Section IV.
	
	For a filter length $M$ and $L$
	reference samples, computing the
	windowed errors and kernel-weighted
	gradient requires $\mathcal{O}(ML)$
	operations per iteration, while the
	elementwise RZA term requires
	$\mathcal{O}(M)$ operations.
	Consequently, the overall
	per-iteration complexity is
	$\mathcal{O}(ML)$, with
	$\mathcal{O}(ML)$ memory required
	to store the reference observations.

	\section{Convergence and Stability Analysis}
	
	This section establishes exact weight-error and
	energy relations for RZA-SSE, followed by local
	mean-stability and approximate steady-state
	MSD analyses.
	
	\subsection{Weight-Error Dynamics}
	
	For a fixed user and BS angular component, let
	$\widetilde{\mathbf{w}}(n)=\mathbf{w}_o-\mathbf{w}(n)$.
	Since all reference errors are evaluated using
	$\mathbf{w}(n)$, the regression model in
	(\ref{eq:adaptive_regression}) yields
	\begin{equation}
		\Delta e_{ni}
		=\widetilde{\mathbf{w}}^H(n)\Delta\mathbf{x}_{ni}
		+\Delta v_{ni},
		\label{eq:analysis_error_difference}
	\end{equation}
	where $\Delta\mathbf{x}_{ni}=\mathbf{x}(n)-\mathbf{x}(i)$
	and $\Delta v_{ni}=v(n)-v(i)$.
	
	With the normalized weights $\pi_{ni}$ defined
	in (\ref{eq:normalized_kernel}), introduce
	\begin{equation}
		\begin{aligned}
			\mathbf{A}_n
			&=\sum_{i=n-L}^{n-1}
			\pi_{ni}\Delta\mathbf{x}_{ni}
			\Delta\mathbf{x}_{ni}^H,\\
			\mathbf{b}_n
			&=\sum_{i=n-L}^{n-1}
			\pi_{ni}\Delta\mathbf{x}_{ni}\Delta v_{ni}^*.
		\end{aligned}
		\label{eq:analysis_quantities}
	\end{equation}
	Also, let
	$\mathbf{r}_n=\operatorname{sgn}(\mathbf{w}(n))
	/(1+\epsilon|\mathbf{w}(n)|)^2$,
	with elementwise operations.
	Since $\pi_{ni}\geq0$ and $\sum_i\pi_{ni}=1$,
	$\mathbf{A}_n$ is Hermitian positive semidefinite.
	
	Substituting (\ref{eq:analysis_error_difference})
	into (\ref{eq:20}) gives the exact recursion
	\begin{equation}
		\widetilde{\mathbf{w}}(n+1)
		={}(\mathbf{I}-\eta\mathbf{A}_n)
		\widetilde{\mathbf{w}}(n)-\eta\mathbf{b}_n+\rho\mathbf{r}_n,
		\label{eq:weight_evolution_rza}
	\end{equation}
	where $\eta=\mu/(2\beta^2)$ and
	$\rho=\mu\gamma/2$. The kernel weights, and
	thus $\mathbf{A}_n$ and $\mathbf{b}_n$,
	depend on the weight error and disturbances.
	
	\subsection{Mean Behavior and Local Stability}
	
	Taking expectations in
	(\ref{eq:weight_evolution_rza}) yields
	\begin{equation}
		\mathbb{E}[\widetilde{\mathbf{w}}(n+1)]
		={}\mathbb{E}[\widetilde{\mathbf{w}}(n)]
		-\eta\mathbb{E}[\mathbf{A}_n
		\widetilde{\mathbf{w}}(n)]-\eta\mathbb{E}[\mathbf{b}_n]
		+\rho\mathbb{E}[\mathbf{r}_n].
		\label{eq:mean_recursion_rza}
	\end{equation}
	This relation is exact. Since $\pi_{ni}$
	depends on the error differences, zero-mean
	noise does not generally imply
	$\mathbb{E}[\mathbf{b}_n]=\mathbf{0}$,
	nor can the first expectation be factorized
	without additional assumptions.
	
	For a simplified local reference model,
	assume
	$\mathbb{E}[\mathbf{A}_n\widetilde{\mathbf{w}}(n)]
	\approx\overline{\mathbf{A}}\,
	\mathbb{E}[\widetilde{\mathbf{w}}(n)]$
	and $\mathbb{E}[\mathbf{b}_n]\approx\mathbf{0}$,
	where $\overline{\mathbf{A}}=\mathbb{E}[\mathbf{A}_n]$
	is approximately constant. If
	$\overline{\mathbf{A}}\succ\mathbf{0}$,
	the homogeneous component is stable when
	\begin{equation}
		0<\mu<
		\frac{4\beta^2}
		{\lambda_{\max}(\overline{\mathbf{A}})}.
		\label{eq:stability_condition}
	\end{equation}
	This condition requires excitation of the
	regressor differences and does not establish
	stability of the complete nonlinear algorithm.
	
	For a local nonlinear analysis, define
	$\mathbf{u}=[\operatorname{Re}\{\mathbf{w}\}^T,
	\operatorname{Im}\{\mathbf{w}\}^T]^T
	\in\mathbb{R}^{2M}$.
	Let $\mathcal{T}(\mathbf{u};\boldsymbol{\xi})$
	denote the RZA-SSE update for a
	training--disturbance window $\boldsymbol{\xi}$.
	Assume a fixed channel and a stationary window
	process; these conditions are not guaranteed
	by finite or deterministic RIS training.
	
	Define the frozen-state mean map
	$\overline{\mathcal{T}}(\mathbf{u})
	=\mathbb{E}_{\boldsymbol{\xi}}
	[\mathcal{T}(\mathbf{u};\boldsymbol{\xi})]$.
	Suppose it admits a fixed point
	$\mathbf{u}_{\star}$ and is continuously
	differentiable nearby, with differentiation
	interchangeable with expectation. Let
	$
	\mathbf{J}_{\star}
	=
	\left.
	\frac{\partial\overline{\mathcal{T}}
		(\mathbf{u})}{\partial\mathbf{u}^{T}}
	\right|_{\mathbf{u}=\mathbf{u}_{\star}}.
	\label{eq:mean_jacobian_rza}
	$
	Then $\varrho(\mathbf{J}_{\star})<1$
	ensures local asymptotic stability of the
	deterministic mean map, not global or
	mean-square stability of the stochastic
	algorithm. The classical Jacobian analysis
	requires a differentiable mean map;
	the RZA attraction itself is nonsmooth
	at zero coefficients, where a separate
	nonsmooth treatment may be necessary.
	The update remains defined with
	$\operatorname{sgn}(0)=0$.
	
	\subsection{Mean-Square Energy Relations}
	
	Let $\|\mathbf{z}\|_{\boldsymbol{\Sigma}}^2
	=\mathbf{z}^H\boldsymbol{\Sigma}\mathbf{z}$
	for Hermitian $\boldsymbol{\Sigma}\succ\mathbf{0}$,
	and define
	$\mathbf{h}_n=\mathbf{A}_n\widetilde{\mathbf{w}}(n)
	+\mathbf{b}_n$ and
	$\mathbf{t}_n=\eta\mathbf{h}_n-\rho\mathbf{r}_n$.
	The exact energy identity is
	\begin{equation}
		\|\widetilde{\mathbf{w}}(n+1)\|_
		{\boldsymbol{\Sigma}}^2
		={}\|\widetilde{\mathbf{w}}(n)\|_
		{\boldsymbol{\Sigma}}^2-2\operatorname{Re}
		\{\widetilde{\mathbf{w}}^H(n)
		\boldsymbol{\Sigma}\mathbf{t}_n\}+\|\mathbf{t}_n\|_{\boldsymbol{\Sigma}}^2.
		\label{eq:weighted_norm_rza}
	\end{equation}
	In particular,
	\begin{equation}
		\mathbf{b}_n^H\boldsymbol{\Sigma}\mathbf{b}_n
		={}\sum_{i,j=n-L}^{n-1}
		\pi_{ni}\pi_{nj}\Delta v_{ni}
		\Delta v_{nj}^*\times\Delta\mathbf{x}_{ni}^H
		\boldsymbol{\Sigma}\Delta\mathbf{x}_{nj}.
		\label{eq:noise_cross_terms_rza}
	\end{equation}
	Off-diagonal terms cannot generally be
	discarded because the differences share
	the current observation and the weights
	are data dependent.
	
	With $\operatorname{MSD}(n)
	=\mathbb{E}\|\widetilde{\mathbf{w}}(n)\|_2^2$,
	setting $\boldsymbol{\Sigma}=\mathbf{I}$
	gives the exact relation
	\begin{equation}
		\begin{aligned}
			&\operatorname{MSD}(n+1)-\operatorname{MSD}(n)=\\
			{}&-2\eta\operatorname{Re}
			\mathbb{E}[\widetilde{\mathbf{w}}^H(n)\mathbf{h}_n]
			+2\rho\operatorname{Re}
			\mathbb{E}[\widetilde{\mathbf{w}}^H(n)\mathbf{r}_n]\\
			&+\eta^2\mathbb{E}\|\mathbf{h}_n\|_2^2
			-2\eta\rho\operatorname{Re}
			\mathbb{E}[\mathbf{h}_n^H\mathbf{r}_n]+\rho^2\mathbb{E}\|\mathbf{r}_n\|_2^2.
		\end{aligned}
		\label{eq:weighted_norm_rza_full}
	\end{equation}
	If a stationary regime with finite second
	moments exists, its MSD increment vanishes.
	Writing $\mathbb{E}_{\infty}$ for stationary
	expectation, the resulting exact balance is
	\begin{equation}
		\begin{aligned}
			2\eta\operatorname{Re}
			\mathbb{E}_{\infty}
			[\widetilde{\mathbf{w}}^H\mathbf{h}]
			={}&\eta^2\mathbb{E}_{\infty}
			\|\mathbf{h}\|_2^2+2\rho\operatorname{Re}
			\mathbb{E}_{\infty}
			[\widetilde{\mathbf{w}}^H\mathbf{r}]\\
			&-2\eta\rho\operatorname{Re}
			\mathbb{E}_{\infty}[\mathbf{h}^H\mathbf{r}]
			+\rho^2\mathbb{E}_{\infty}
			\|\mathbf{r}\|_2^2.
		\end{aligned}
		\label{eq:steady_state_rza}
	\end{equation}
	This balance is implicit because its terms
	contain coupled nonlinear random variables.
	
	\subsection{Approximate Steady-State MSD}
	
	Consider a differentiable mean-field fixed
	point $\mathbf{u}_{\star}$ and assume that
	the iterates remain in a neighborhood where
	higher-order terms can be neglected. Define
	$\boldsymbol{\delta}(n)=\mathbf{u}(n)-\mathbf{u}_{\star}$
	and the equilibrium fluctuation
	$\boldsymbol{\zeta}_n=
	\mathcal{T}(\mathbf{u}_{\star};
	\boldsymbol{\xi}_n)-\mathbf{u}_{\star}$.
	By the fixed-point condition,
	$\mathbb{E}[\boldsymbol{\zeta}_n]=\mathbf{0}$.
	
	Replacing the stochastic update Jacobian
	by $\mathbf{J}_{\star}$ and neglecting
	multiplicative-noise terms gives
	\begin{equation}
		\boldsymbol{\delta}(n+1)
		\simeq\mathbf{J}_{\star}
		\boldsymbol{\delta}(n)+\boldsymbol{\zeta}_n.
		\label{eq:local_error_rza}
	\end{equation}
	For this linear approximation, treat
	$\boldsymbol{\zeta}_n$ as a stationary
	exogenous input with lag covariance
	$\mathbf{C}_k=\mathbb{E}[
	\boldsymbol{\zeta}_n
	\boldsymbol{\zeta}_{n-k}^{T}]$, $k\geq0$.
	Assume $\varrho(\mathbf{J}_{\star})<1$
	and absolutely summable lag covariances.
	
	The stationary covariance $\mathbf{P}$
	of the linear approximation satisfies
	\begin{equation}
		\mathbf{P}\simeq
		\mathbf{J}_{\star}\mathbf{P}
		\mathbf{J}_{\star}^{T}
		+\mathbf{Q}_{\mathrm{eff}},
		\label{eq:lyapunov_rza}
	\end{equation}
	where
	$\mathbf{Q}_{\mathrm{eff}}
	={}\mathbf{C}_0+
	\sum_{k=1}^{\infty}
	\big[
	\mathbf{J}_{\star}^{k}\mathbf{C}_k^T
	+\mathbf{C}_k(\mathbf{J}_{\star}^{T})^{k}
	\big].$
	
	The effective forcing matrix need not be positive
	semidefinite because it incorporates temporal
	cross-covariance contributions. If the underlying
	training--disturbance samples are independent and
	identically distributed, the frozen-state innovation process depends on overlapping finite observation windows. Consequently, its lag covariances vanish
	for $k>L$, provided that each innovation depends
	only on the current sample and the preceding $L$
	samples. This finite-memory property does not
	necessarily hold for temporally correlated
	disturbances or structured RIS training.
	For temporally uncorrelated innovations,
	$\mathbf{Q}_{\mathrm{eff}}=\mathbf{C}_0$.
	
	Vectorization gives
	\begin{equation}
		\operatorname{vec}(\mathbf{P})
		\simeq
		[\mathbf{I}_{4M^2}
		-\mathbf{J}_{\star}\otimes\mathbf{J}_{\star}]^{-1}
		\operatorname{vec}(\mathbf{Q}_{\mathrm{eff}}).
		\label{eq:msd_covariance_solution}
	\end{equation}
	Let $\mathbf{u}_o=
	[\operatorname{Re}\{\mathbf{w}_o\}^T,
	\operatorname{Im}\{\mathbf{w}_o\}^T]^T$.
	The resulting approximate steady-state MSD is
	\begin{equation}
		{
			\operatorname{MSD}_{\infty}^{\mathrm{RZA\text{-}SSE}}
			\simeq
			\|\mathbf{u}_o-\mathbf{u}_{\star}\|_2^2
			+\operatorname{Tr}(\mathbf{P})
		}
		\label{eq:finalMSD_rza}
	\end{equation}
	which separates the squared equilibrium bias
	from the local fluctuation variance.
	
	For unregularized SSE ($\rho=0$), the same
	approximation yields
	\begin{equation}
		\operatorname{MSD}_{\infty}^{\mathrm{SSE}}
		\simeq
		\|\mathbf{u}_o-\mathbf{u}_{\star,0}\|_2^2
		+\operatorname{Tr}(\mathbf{P}_0),
		\label{eq:finalMSD_unreg}
	\end{equation}
	where $\mathbf{u}_{\star,0}$ and $\mathbf{P}_0$
	are the corresponding mean-field equilibrium
	and local covariance. Hence, RZA changes both
	the equilibrium bias and fluctuation covariance,
	rather than adding an isolated $\rho^2$ penalty.
	
	This characterization is local and approximate.
	Its numerical evaluation requires the mean-field
	equilibrium, Jacobian, and innovation covariances
	under the specified RIS training and GMM noise
	models; predictive accuracy must be verified
	through Monte Carlo simulations.

	\section{Simulation Results}
	\label{sec:simulation}
	
	This section evaluates the performance of the proposed
	RZA-SSE algorithm for angular-domain CE
	in RIS-assisted mmWave massive MIMO systems under
	Gaussian-mixture impulsive noise. The proposed estimator
	is compared with the NLMS-NLMF-LOG algorithm
	\cite{shao2025adaptive} and the RZA-MCC algorithm
	\cite{hajiabadi}. The performance metric is the
	MSD, defined as
	$
	\mathrm{MSD}(n)
	=
	\mathbb{E}
	\left[
	\left\|
	\mathbf{w}_{o}-\mathbf{w}(n)
	\right\|_2^2
	\right],
	$
	which is expressed in decibels as
	$10\log_{10}\mathrm{MSD}(n)$.
	The numerical results examine convergence behavior
	and steady-state estimation accuracy as functions of
	the adaptation step size, SNR,
	and kernel bandwidth.
	
	\subsection{Simulation Setup}
	
	We consider an uplink RIS-assisted mmWave massive
	MIMO system comprising a base station equipped with
	$N_B=64$ antennas, an RIS with $N_R=256$
	reflecting elements, and $K=16$ single-antenna users.
	The propagation channels follow the geometric
	Saleh--Valenzuela model described in Section II,
	with $L_b=5$ dominant RIS--BS paths and
	$L_{r,k}=8$ dominant user--RIS paths.
	The angular-domain channel representation
	is assumed to be sparse, consistent with
	the limited number of dominant propagation
	paths in the adopted geometric channel model.
	For each estimated angular component, the adaptive
	filter length is $M=N_R=256$.
	The received observations are corrupted by
	non-Gaussian impulsive disturbances represented
	by the complex Gaussian-mixture model introduced
	in Section II. The sensing vectors are associated
	with the RIS training configurations through
	$\mathbf{x}(n)=[\mathbf{U}_R^H
	\boldsymbol{\Theta}]_{:,n}$, where the RIS
	reflection coefficients satisfy the unit-modulus
	constraint.
	Unless otherwise stated, the RZA-SSE parameters
	are set to $\beta=3$, $L=35$,
	$\rho=10^{-6}$, and $\epsilon=5$.
	The adaptation step size is $\mu=0.03$
	except in the step-size experiment.
	According to the update derived in Section III,
	the effective gradient coefficient is
	$\eta=\mu/(2\beta^2)$, which gives
	$\eta=1.667\times10^{-3}$ for
	$\mu=0.03$ and $\beta=3$.
	An SNR of $12$ dB is employed for the
	convergence, step-size, and kernel-bandwidth
	experiments unless otherwise specified.

	\subsection{Performance Evaluation}
	
\begin{figure}[!t]
		\centering
		\includegraphics[width=\columnwidth]{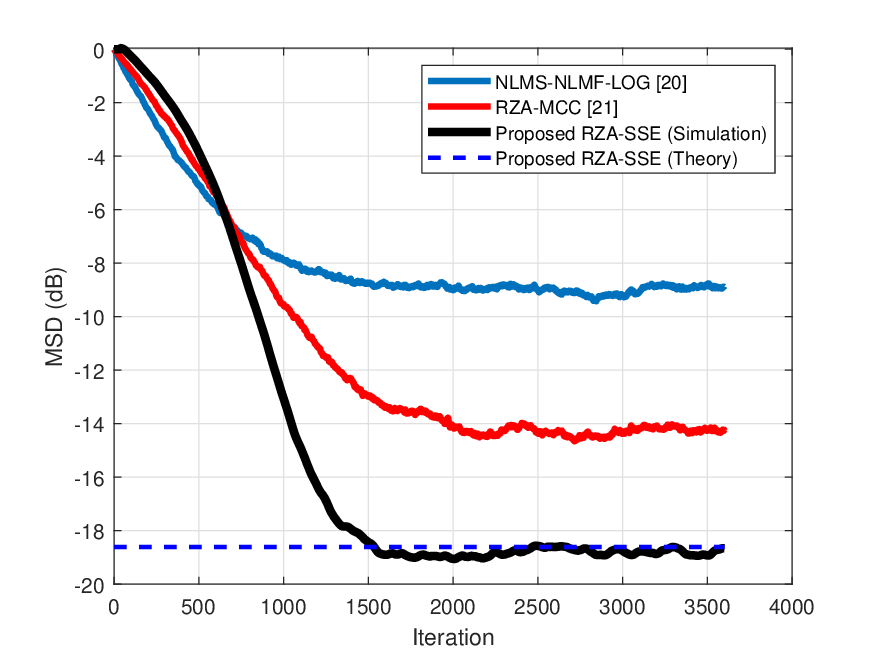}
		\caption{MSD learning curves of the proposed
			RZA-SSE estimator, RZA-MCC, and NLMS-NLMF-LOG
			at an SNR of $12$ dB.}
		\label{fig:convergence}
	\end{figure}

	Fig.~\ref{fig:convergence} compares the MSD learning
	curves of the proposed RZA-SSE estimator with those
	of RZA-MCC and NLMS-NLMF-LOG at an SNR of
	$12$ dB. RZA-SSE exhibits a faster reduction in MSD
	and reaches its steady-state region after approximately
	$1500$ iterations, whereas RZA-MCC requires about
	$2000$ iterations. NLMS-NLMF-LOG exhibits a
	substantially higher residual MSD throughout the
	steady-state region.
	The steady-state MSD values are approximately
	$-18.8$, $-14.2$, and $-9.0$ dB for RZA-SSE,
	RZA-MCC, and NLMS-NLMF-LOG, respectively.
	Accordingly, RZA-SSE achieves an MSD improvement
	of approximately $4.6$ dB over RZA-MCC and
	$9.8$ dB over NLMS-NLMF-LOG. These results
	demonstrate the favorable convergence behavior
	and steady-state estimation accuracy of RZA-SSE
	under the considered impulsive-noise conditions.
	
\begin{figure}[!t]
		\centering
		\includegraphics[width=\columnwidth]{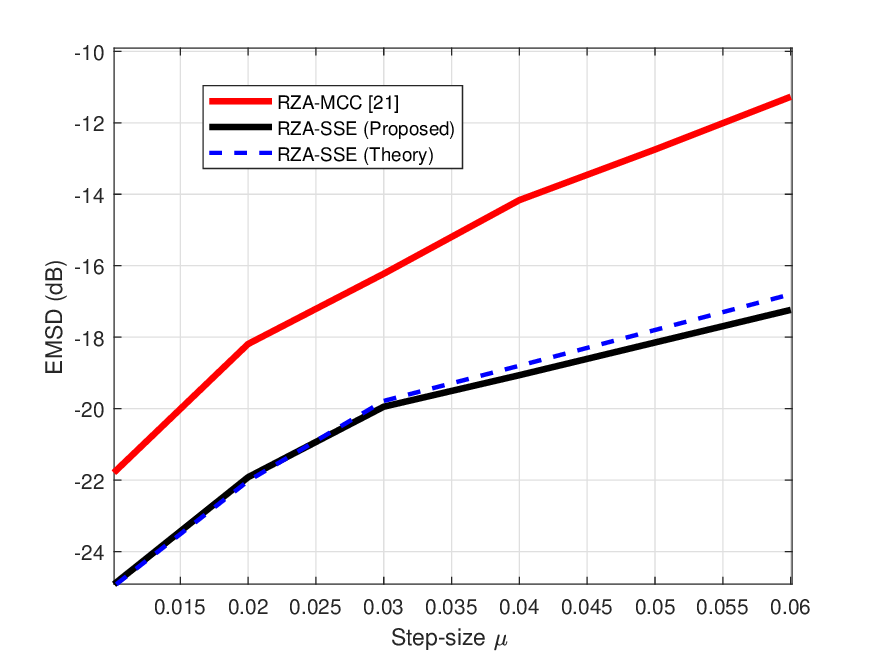}
		\caption{Steady-state MSD versus adaptation
			step size $\mu$ for RZA-SSE and RZA-MCC
			at an SNR of $12$ dB.}
		\label{fig:stepsize}
	\end{figure}

	Fig.~\ref{fig:stepsize} compares the steady-state
	MSD of RZA-SSE and RZA-MCC as a function of
	the adaptation step size $\mu$ over the range
	$0.01 \leq \mu \leq 0.06$ at an SNR of $12$ dB.
	RZA-SSE consistently achieves a lower MSD than
	RZA-MCC throughout the investigated range.
	Specifically, the MSD of RZA-SSE increases
	from approximately $-25$ dB at $\mu=0.01$
	to $-17.5$ dB at $\mu=0.06$, whereas RZA-MCC
	exhibits a similar trend at higher MSD levels.
	The observed increase in steady-state MSD
	with $\mu$ is consistent with the
	misadjustment behavior of stochastic-gradient
	adaptive algorithms. For RZA-SSE, the effective
	gradient coefficient is $\eta=\mu/(2\beta^2)$,
	while the reweighted attraction coefficient
	satisfies $\rho=\mu\gamma/2$. Consequently,
	the step-size dependence is influenced by
	both the adaptation gain and the adopted
	regularization parameterization. The results
	illustrate the sensitivity of RZA-SSE to
	the step size within the evaluated range,
	without implying stability beyond it.
	
\begin{figure}[!t]
		\centering
		\includegraphics[width=\columnwidth]{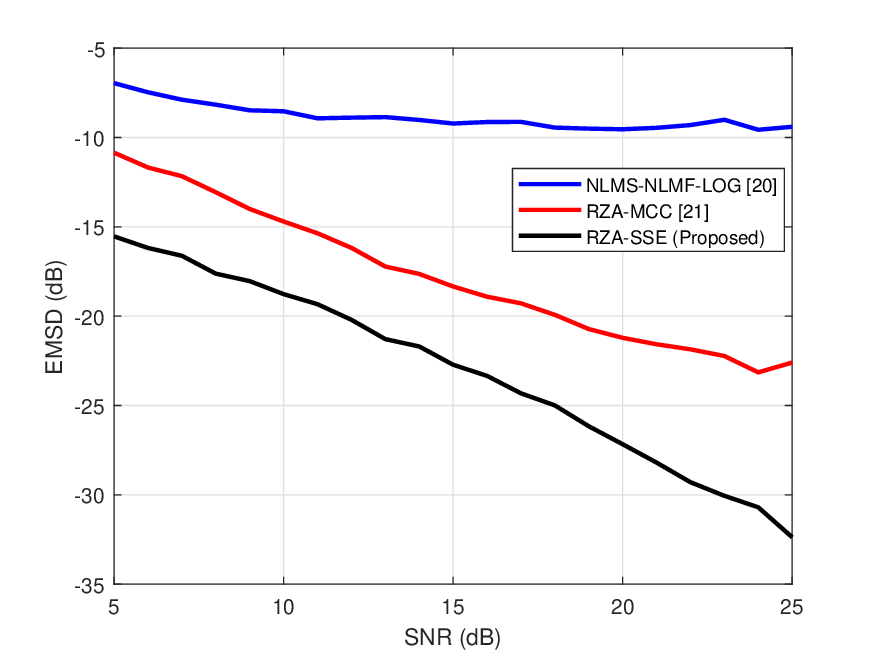}
		\caption{Steady-state MSD versus SNR for
			the proposed RZA-SSE estimator,
			RZA-MCC, and NLMS-NLMF-LOG
			at $\mu=0.03$.}
		\label{fig:snr}
	\end{figure}

	Fig.~\ref{fig:snr} compares the steady-state MSD
	of RZA-SSE, RZA-MCC, and NLMS-NLMF-LOG over
	an SNR range of $5$--$25$ dB, with
	$\mu=0.03$. RZA-SSE consistently achieves
	the lowest MSD throughout the investigated
	range. Specifically, its MSD decreases from
	approximately $-15.5$ dB at an SNR of $5$ dB
	to $-32$ dB at $25$ dB. Although RZA-MCC
	also exhibits improved estimation accuracy
	with increasing SNR, NLMS-NLMF-LOG shows
	a comparatively limited improvement, with
	MSD values remaining between approximately
	$-7$ and $-10$ dB.
	The results indicate that RZA-SSE maintains
	a performance advantage over both benchmark
	algorithms across the considered noise levels.
	The separation between the MSD curves varies
	with SNR, suggesting that the relative benefit
	of the proposed estimator depends on the
	operating conditions. This behavior is
	consistent with the combined use of
	kernel-based error adaptation and reweighted
	sparsity regularization under the simulated
	Gaussian-mixture impulsive noise.
	
\begin{figure}[!t]
		\centering
		\includegraphics[width=\columnwidth]{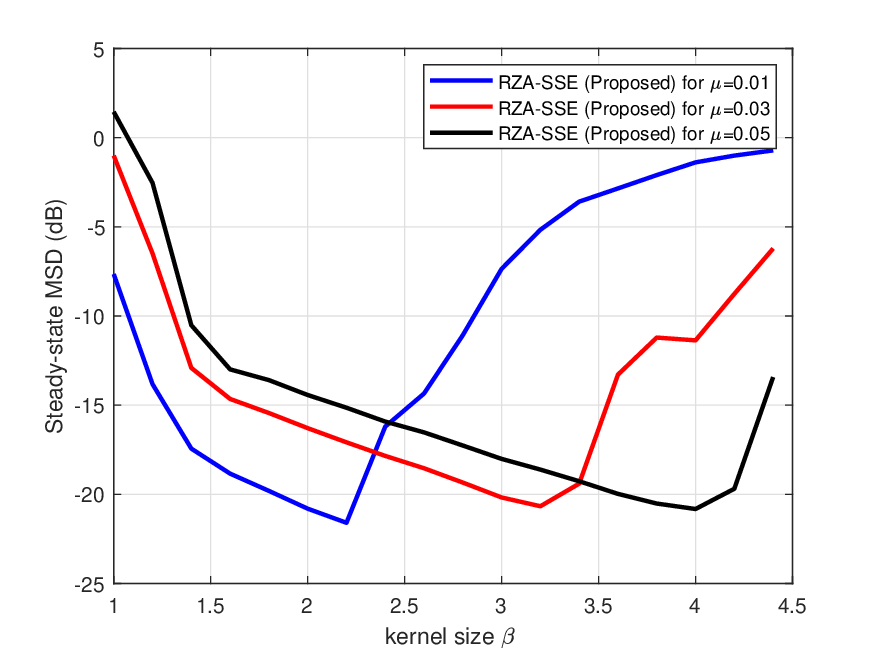}
		\caption{Steady-state MSD of the proposed
			RZA-SSE estimator versus Gaussian kernel
			bandwidth $\beta$ at an SNR of $12$ dB
			and different step-sizes $\mu=0.01, 0.03, 0.05$.}
		\label{fig:kernelsize}
	\end{figure}

	Fig.~\ref{fig:kernelsize} illustrates the
	effect of the Gaussian kernel bandwidth
	$\beta$ on the steady-state MSD of RZA-SSE
	at an SNR of $12$ dB and $\mu=0.03$.
	The MSD exhibits a nonmonotonic dependence
	on $\beta$, initially decreasing as the
	bandwidth increases and reaching a minimum
	of approximately $-20.7$ dB near
	$\beta=3.2$. Beyond this value, further
	increases in $\beta$ lead to higher
	steady-state MSD, indicating the existence
	of a favorable bandwidth region for the
	considered configuration.
	The observed behavior reflects the combined
	influence of $\beta$ on the normalized
	kernel weights and the effective adaptation
	gain $\eta=\mu/(2\beta^2)$. Consequently,
	varying $\beta$ at a fixed $\mu$ changes
	both the relative weighting of error
	differences and the gradient scaling.
	These results highlight the importance
	of selecting an appropriate kernel bandwidth
	to achieve favorable steady-state estimation
	accuracy under the considered impulsive-noise
	conditions.
	
	Overall, the numerical results indicate a
	consistent steady-state MSD advantage for
	RZA-SSE over the selected benchmarks under
	the evaluated conditions. Additional ablation
	experiments are needed to isolate the
	contributions of kernel-based adaptation and
	reweighted sparsity regularization. Independent
	numerical evaluation of the local steady-state
	MSD approximation remains necessary to assess
	the quantitative accuracy of the theoretical
	analysis.

	\section{Conclusion}
	
	This paper proposed RZA-SSE, a sparsity-aware
	adaptive channel estimator for RIS-assisted
	mmWave massive MIMO systems under
	Gaussian-mixture impulsive noise.
	The proposed algorithm combines a stochastic
	Shannon-entropy-based error criterion with
	reweighted zero-attraction regularization
	to improve robustness against large error
	differences while exploiting angular-domain
	channel sparsity.
	The complex-valued update was derived using
	Wirtinger calculus, and exact weight-error
	and weighted energy recursions were established.
	Mean stability and steady-state MSD were
	investigated under explicitly stated local
	approximations and assumptions.
	Simulation results showed that RZA-SSE
	achieved lower steady-state MSD than RZA-MCC
	and NLMS-NLMF-LOG across the evaluated
	operating conditions, with faster convergence
	in the reported learning-curve experiment.
	The results also demonstrated the dependence
	of estimation accuracy on the step size,
	SNR, and kernel bandwidth.
	Future work will consider time-varying
	channels, structured RIS training, and
	near-field propagation scenarios.

\end{document}